4# Perfect Domain-Lattice Matching Between MgB$_2$ and Al$_2$O$_3$: Single-Crystal MgB$_2$ Thin Films Grown on Sapphire

Won Kyung Seong,[1] Sangjun Oh,[2] and Won Nam Kang[1,*]

[1]BK21 Physics Division and Department of Physics, Sungkyunkwan University, Suwon 440-746, Republic of Korea.

[2]National Fusion Research Institute (NFRI), Daejeon 305-806, Republic of Korea.We have found that single-crystal films can be grown on (0001) Al$_2$O$_3$ substrates through the golden relation of a perfect lattice-matching ratio ($8:3\sqrt{3}$) between the *a*-axis lattice constants of MgB$_2$ and Al$_2$O$_3$. Selected area electron diffraction patterns evidently indicate hexagonal MgB$_2$ film with a 30$^o$ in-plane rotation with respect to the Al$_2$O$_3$ substrate. The films grown on Al$_2$O$_3$ show a zero-resistance transition temperature of 39.6 K with a transition width of 0.1 K. The critical current density at zero field is comparable to the depairing critical current density and rapidly decreases with increasing applied field due to the lack of pinning sites, as observed for high-quality MgB$_2$ single crystals.PACS numbers: 68.55.at, 68.37.Og, 74.78.Fk, 74.70.Ad

*Electronic address: wnkang@skku.edu

Since the discovery of superconductivity in $MgB_2$ with a transition temperature ($T_c$) close to 40 K, extensive studies, both theoretical and experimental, have been carried out [1-4]. $MgB_2$ is a two-gap superconductor, which was confirmed by the existence of four disconnected sheets of the Fermi surface: two hole-like $\sigma$ bands, a hole- and an electron-like $\pi$ bands [5,6]. The two-band nature not only influences various superconducting properties of $MgB_2$ but also causes new physical phenomena [7-9], and can be modified by chemical doping or substitution [10-13]. Currently, most fundamental research on $MgB_2$ is carried out using single crystals, requires both high-pressure and high-temperature fabrication conditions, and takes a very long growth time [14,15]. At the moment, however, the sizes of the $MgB_2$ single crystals are very small [14]. As a result, $MgB_2$ thin films have been produced and employed in the study of two-gap superconductivity in $MgB_2$. Furthermore, strongly linked nature of the grain connectivity [2] and a high charge carrier density [4,16] are very suitable for engineering applications, such as electronic devices [17] and radio-frequency (RF) cavities [18,19]. For example, Nb-based RF cavity technology is now very close to theoretical limits [18] and high-quality single-crystalline $MgB_2$ thin films are considered to be the most plausible candidates [19] for meeting the still expanding requirements of accelerator research.

However, the potential applications of $MgB_2$ thin films have not yet been realized because technique for fabricating high-quality single-crystalline films has not been developed. Fabrication of a single-crystalline film on a cheap substrate, such as sapphire ($Al_2O_3$), will be more beneficial for practical applications. The excellent thermal conductivity and the optical properties of $Al_2O_3$ provide further advantages for practical applications at low temperatures. If we consider the large lattice mismatch of 23% between MgB$_2$ and Al$_2$O$_3$, growth of single-crystal MgB$_2$ films on Al$_2$O$_3$ is a very challenging task.

In this Letter, we report the fabrication of single-crystal MgB$_2$ thin films with a large area on *c*-plane Al$_2$O$_3$ substrates by using HPCVD (hybrid physical-chemical vapor deposition). We have found that single-crystal films can be grown on Al$_2$O$_3$ substrates with a lattice-matching ratio of $8:3\sqrt{3}$ between the *a*-axis lattice constants of MgB$_2$ and Al$_2$O$_3$ where the $[11\bar{2}0]$ directions are rotated by 30$^\mathrm{o}$ with respect to each other.

The HPCVD process is well known to be a very promising technique for the fabrication of high-quality MgB$_2$ films and to be suitable for mass production [3,20-22]. The main process chamber of the HPCVD system is connected to a load-lock chamber on the bottom, and an induction heater is attached on the outer periphery of the process chamber. First, the process chamber is purged with ultra-high-purity (99.9999%) Ar gas. A sapphire substrate and granular-shaped Mg pieces (6.5 g) are placed on a susceptor. The susceptor is then moved to the process chamber from the load-lock chamber. The substrate can be inductively heated to 700 $^\mathrm{o}$C. The usual heating rate is 7 $^\mathrm{o}$C min$^{-1}$, and 200 Torr of pure H$_2$ (99.9999%) gas is introduced. After the temperature has been stabilized, B$_2$H$_6$ (5% B$_2$H$_6$ in H$_2$) gas is supplied to initiate film deposition. The total flow rate of the mixed gas is 150 sccm. After the deposition, the susceptor is cooled to room temperature by flowing pure H$_2$ gas.

The surface morphologies of the MgB$_2$ thin films were examined by using scanning electron microscopy (SEM). As can be seen in Figs. 1(a) and (b), the surface morphology of the MgB$_2$ thin films was very smooth without grain boundaries. The root-mean-square surface roughness was as low as 1.5 nm over a wide area of 3 μm × 3 μm, much smoother than any other epitaxial $MgB_2$ thin films reported so far [3,22,23]. Four-circle X-ray diffraction (XRD) results are shown in Figs. 1(c) and (d). Only (000$l$) peaks, without any impurity or secondary phase peaks, are observed. The $c$-axis lattice constant determined from the (0002) peak is 3.519 Å, slightly smaller than the bulk value of 3.521 Å. The full width at half maximum (FWHM) obtained from the rocking curve for the (0002) peak (inset of Fig. 1(c)) is as low as 0.13°, which is very close to that for a $MgB_2$ single crystal [24] and is much narrower than that for an epitaxial $MgB_2$ thin film grown on a (0001) SiC substrate [25], indicating that the $MgB_2$ thin film grown on a (0001) $Al_2O_3$ substrate has a single-crystal quality.

Bright-field transmission electron microscope (TEM) images (Fig. 2(a): cross-sectional view, (b): planar view) also indicate a single-crystal quality without grain boundaries for the samples prepared in this study. TEM specimens were prepared by using a lift-out technique with a dual-beam focused ion beam. The $MgB_2$ thin film with a thickness 780 nm was very smooth without grain boundaries, as shown in Figs. 2(a) and (b). Selected-area electron diffraction (SAED) patterns (Figs. 2(c) and (d)) were obtained to characterize the growth mechanism of the single-crystal $MgB_2$ film on a (0001) $Al_2O_3$ substrate. The SAED patterns for the interface between the $MgB_2$ film and the (0001) $Al_2O_3$ substrate are shown in Fig 2(c). Very clear spots can be classified into two sets: one corresponds to the $MgB_2$ film, the other belongs to the substrate. No other phase diffraction spots can be observed, consistent with the XRD results. This is further evidence that the $MgB_2$ thin film grown on the (0001) $Al_2O_3$ substrate is a high-quality single crystal. The $MgB_2$ [000$l$] direction is parallel to the $Al_2O_3$ [000$l$] direction (Fig. 2(c)), indicating that the $c$-axis of the MgB$_2$ thin film is oriented perpendicular to the substrate surface. On the other hand, the [10$\bar{1}$0] direction of the MgB$_2$ film is parallel to the [11$\bar{2}$0] direction of the Al$_2$O$_3$ substrate (Fig. 2(c)), which definitely shows that the *a*-axis of the MgB$_2$ film is rotated by 30$^o$ with respect to that of the Al$_2$O$_3$ substrate.

The SAED patterns for the surface of the MgB$_2$ film (Fig. 2(d)) clearly reflect the hexagonal structure of the single-crystal MgB$_2$. A six-fold symmetry over a large area has also been found in the $\phi$-scan spectrum for the (10$\bar{1}$1) MgB$_2$ plane [3,20,22] (Fig. 1(d)). The planar hexagonal structure can be more directly seen in the high-resolution TEM image shown in Fig. 2(e). All these results suggest that the MgB$_2$ thin film is well aligned not only along the *c*-axis but also along the planar direction. The *a*-axis lattice parameter was found to be 3.088 Å, which is slightly larger than that of 3.086 Å reported [26,27] for bulk MgB$_2$.

The lattice mismatch between MgB$_2$ and Al$_2$O$_3$ is as large as 23% where the *a*-axis lattice constant of Al$_2$O$_3$ is 4.758 Å [20]. A question arises as to how a single-crystal MgB$_2$ thin film can grow so perfectly over on a (0001) Al$_2$O$_3$ substrate. A clue lies in the cross-sectional SAED patterns that the [10$\bar{1}$0] direction of the MgB$_2$ layer is aligned parallel to the [11$\bar{2}$0] direction of the Al$_2$O$_3$ substrate. A perspective schematic illustration at the interface between MgB$_2$ and Al$_2$O$_3$ is presented in Fig. 2(f), where the hexagonal Al$_2$O$_3$ and MgB$_2$ lattices are depicted as blue and red colors, respectively. The *a*-axis of the MgB$_2$ film is rotated 30$^o$ with respect to that of the substrate, as highlighted with yellow colors in Fig. 2(f). This 30$^o$ rotation leads to an interesting long-range lattice-matching order.

The additional points where the locations of Mg and Al coincide with each other are colored green in Fig. 2(f). The growth mechanism exhibited a perfect domain-lattice-matching ratio of $8a_{Mg} : 3\sqrt{3}a_{Al}$ between the *a*-axis lattice constant ($a_{Mg}$) of MgB$_2$ and the *a*-axis lattice constant ($a_{Al}$) of Al$_2$O$_3$ for the case in which the (11$\bar{2}$0) plane of MgB$_2$ was parallel to the (10$\bar{1}$0) plane of Al$_2$O$_3$. Indeed, $8a_{Mg}$ = 24.704 Å and $3\sqrt{3}a_{Al}$ = 24.727 Å, showing a perfect matching relation with a mismatching ratio of less than 0.1%. These findings are our main result; as a result, we are able to successfully grow single-crystal MgB$_2$ thin films with superior quality on (0001) Al$_2$O$_3$ substrates.

As noted, the single-crystal MgB$_2$ thin films have a slightly shorter *c*-axis parameter and a slightly longer *a*-axis parameter compared with the bulk values. Lattice distortions are proposed as a cause for the enhancement of the $T_c$ due to a tensile strain [21,22]. The onset transition temperature of the single-crystal MgB$_2$ thin film is observed to be 40 K with a very sharp transition width of 0.1 K (inset of Fig. 3(a)), which is about 1 K higher than that of bulk MgB$_2$ and seems to be related with the variation in the tensile strain. The temperature dependence of the resistivity of the MgB$_2$ thin film is shown in Fig. 3(a). The resistivity at room temperature (300 K) was about 6.7 μΩ·cm, but it decreased to 0.24 μΩ·cm at 41 K. The residual resistivity ratio (RRR) of the MgB$_2$ thin film grown on (0001) Al$_2$O$_3$ is as high as 28. The high RRR value and the very low resistivity at 41 K indicate that the MgB$_2$ film has a very long electron mean free path due to the low number of crystallographic defects [25].

The single-crystal $MgB_2$ films also showed excellent current-carrying capability. The field dependences of the critical current density ($J_c$) at 5 K and 20 K for single-crystal $MgB_2$ thin films are shown in Fig. 3(b). The critical current density was calculated from the magnetization (*M-H*) hysteresis loops by using the Bean's critical state model [2,4,20]. The self-field critical current density was as high as $5 \times 10^7$ A cm$^{-2}$ at 5 K and $3 \times 10^7$ A cm$^{-2}$ at 20 K, which is very close to the depairing current density [28] and shows the ultimate current-carrying capability [25] of $MgB_2$. The critical current density rapidly decreased with an increasing applied field due to the lack of pinning centers in clean films [26,27], as observed for single crystals.

In summary, we have shown that single-crystal $MgB_2$ thin films can be fabricated on an inexpensive sapphire substrate by using the HPCVD process. The epitaxial growth mechanism on (0001) $Al_2O_3$ is found to be related with the 'golden' long-range lattice-matching order where the *a*-axis of the $MgB_2$ layer is rotated 30$^\circ$ with respect to that of the substrate. As-grown $MgB_2$ thin films on (0001) $Al_2O_3$ were confirmed to have a single-crystal quality with a very smooth surface morphology and excellent superconducting properties, and the self-field critical current density of the as-grown thin film is close to the depairing current density of $MgB_2$. These results will provide a new horizon in superconducting device applications, such as radio-frequency cavities for particle accelerators and Josephson junction devices.

This work was supported by the Mid-career Researcher Program through a National Research Foundation (NRF) grant funded by the Ministry of Education, Science, and

**Figure and Figure captions**

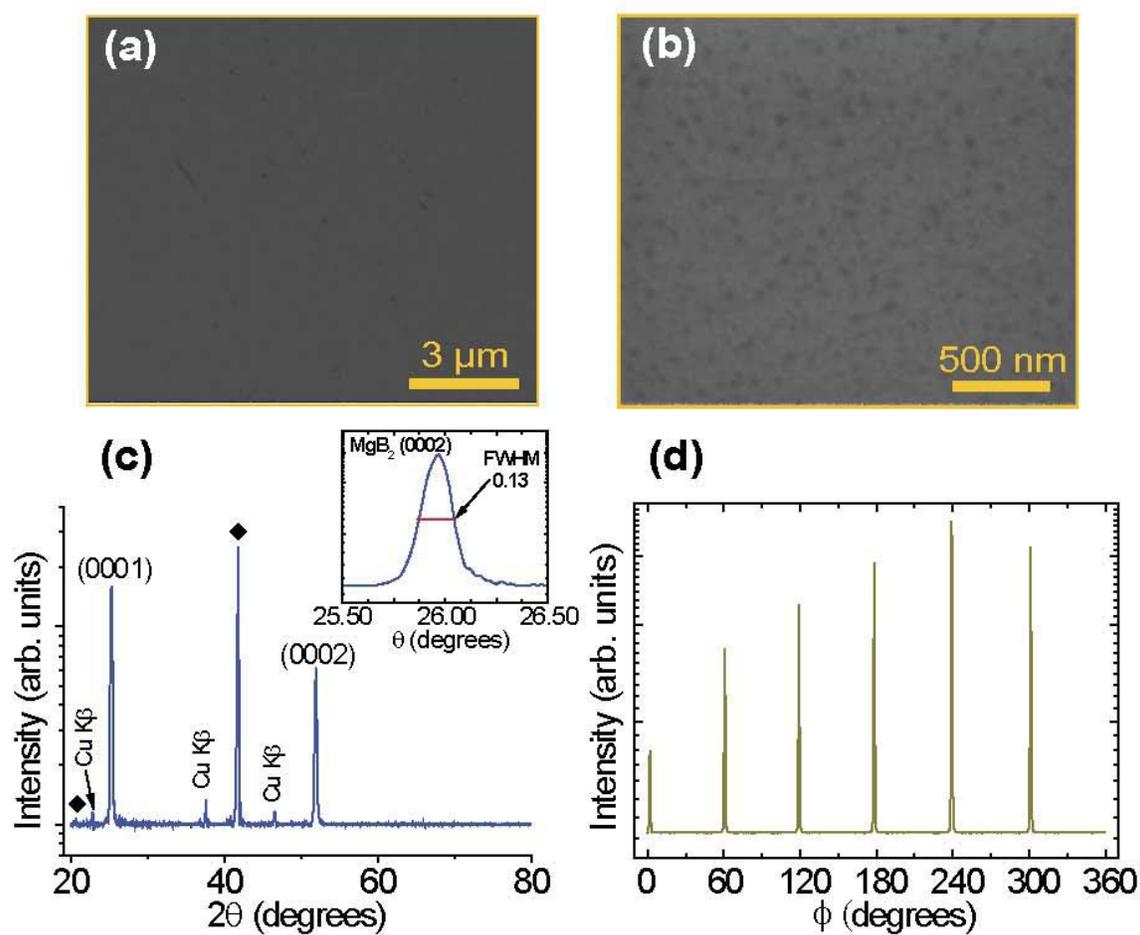

FIG. 1 (color). Surface morphology and XRD measurement results of an as-grown single-crystal MgB$_2$ film on a (0001) Al$_2$O$_3$ substrate: (a) low-magnification SEM image, (b) high-magnification SEM image, (c) $\theta$-$2\theta$ XRD scan, showing only the (*000l*) peaks of MgB$_2$, with substrate peaks being marked by '♦', and (d) $\phi$-scan for (10$\bar{1}$1) MgB$_2$.

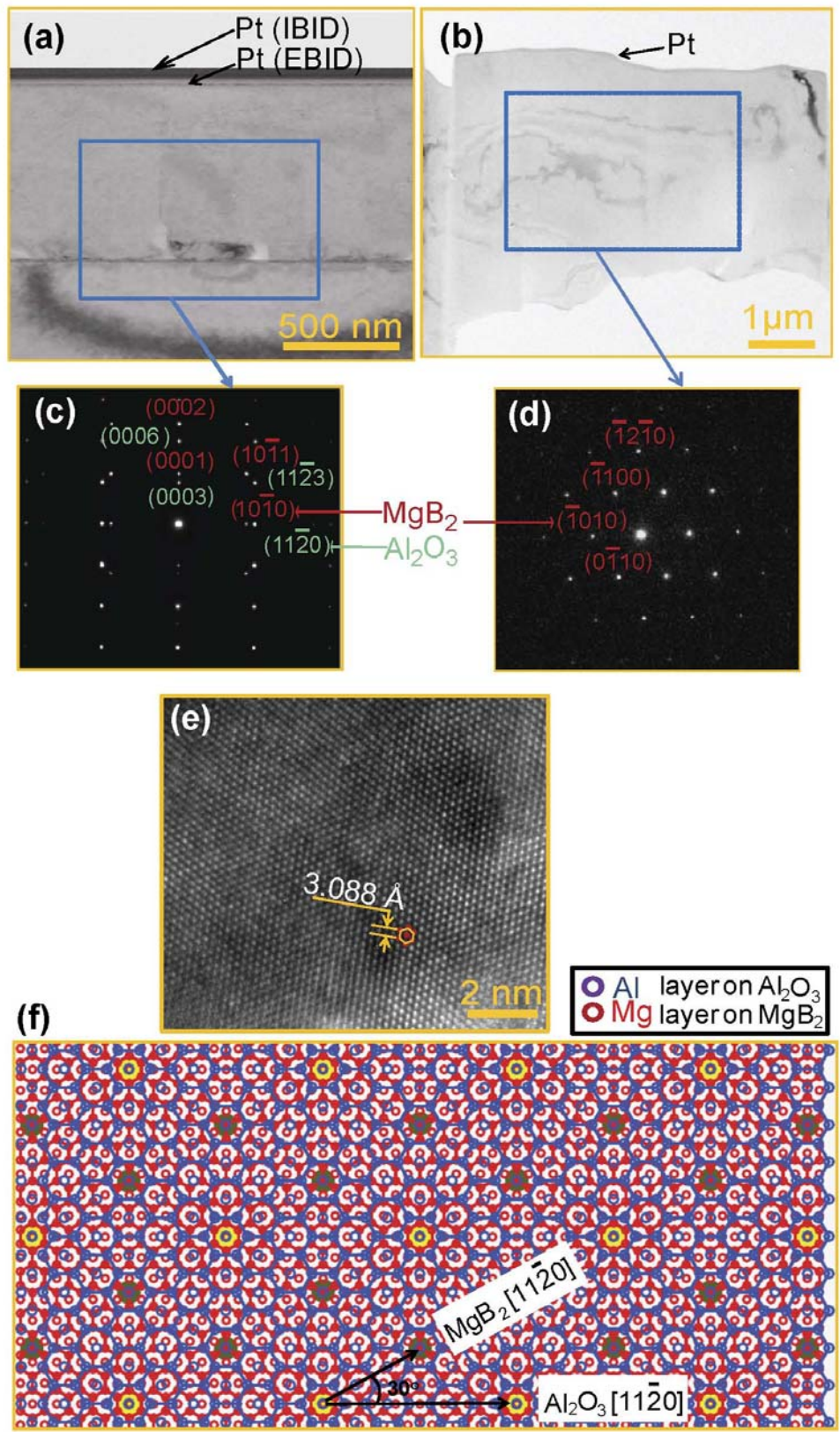

FIG. 2 (color). TEM, selected-area electron diffraction (SAED) investigation and an illustration for epitaxial growth. (a) A low-magnification bright-field cross-sectional TEM image. Platinum was deposited for protection. (b) A low-magnification bright-field planar TEM image. (c) A cross-sectional selected-area electron diffraction (SAED) pattern near the interface region. (d) A planar SAED pattern from the top surface of the $MgB_2$ layer. (e) A high-resolution TEM image. (f) A perspective schematic illustration for the observed perfect long-range lattice-matching relation between the $Al_2O_3$ lattice (blue color) and $MgB_2$ (red color). The $a$-axis of the $MgB_2$ layer is rotated $30^\circ$ with respect to that of the substrate (yellow color). A long-range order can be seen at the green-color circles.

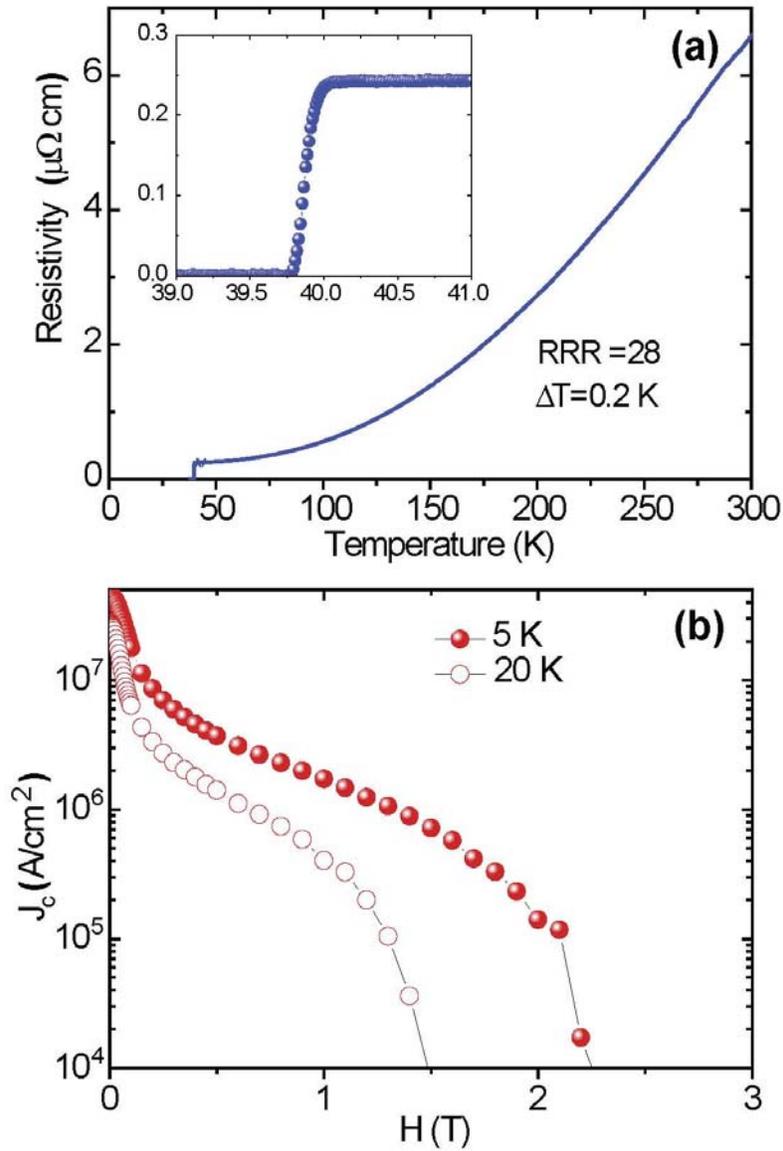

FIG. 3 (color). Superconductivity of a single-crystal $MgB_2$ thin film: (a) resistivity versus temperature for an $MgB_2$ thin film on an $Al_2O_3$ substrate and (b) magnetic field dependence of the critical current density. A rapid increase of the critical current density with increasing field can be observed due to the lack of pinning sites in single-crystal $MgB_2$ films.